# From sequence to protein structure and conformational dynamics with AI/ML

Alexander M. Ille[1,*], Emily Anas[2], Michael B. Mathews[3,4], and Stephen K. Burley[5,6,7,8,9]

[1]Rutgers Cancer Institute, Rutgers, The State University of New Jersey, Newark, NJ
[2]College of Computing, Georgia Institute of Technology, Atlanta, GA
[3]Department of Medicine, Rutgers New Jersey Medical School, Newark, NJ
[4]School of Graduate Studies, Rutgers, The State University of New Jersey, Newark, NJ
[5]Research Collaboratory for Structural Bioinformatics Protein Data Bank, Institute for Quantitative Biomedicine, Rutgers, The State University of New Jersey, Piscataway, NJ
[6]Department of Chemistry and Chemical Biology, Rutgers, The State University of New Jersey, Piscataway, NJ
[7]Rutgers Data Science and Artificial Intelligence (RAD) Collaboratory, Rutgers, The State University of New Jersey, Piscataway, NJ
[8]Rutgers Cancer Institute, Rutgers, The State University of New Jersey, New Brunswick, NJ
[9]Research Collaboratory for Structural Bioinformatics Protein Data Bank, San Diego Supercomputer Center, University of California-San Diego, La Jolla, San Diego, CA

*Correspondence: mai86@rutgers.edu

**Abstract**

The 2024 Nobel Prize in Chemistry was awarded in part for protein structure prediction using AlphaFold2, an artificial intelligence/machine learning (AI/ML) model trained on vast amounts of sequence and 3D structure data. AlphaFold2 and related models, including RoseTTAFold and ESMFold, employ specialized neural network architectures driven by attention mechanisms to infer relationships between sequence and structure. At a fundamental level, these AI/ML models operate on the long-standing hypothesis that the structure of a protein is determined by its amino acid sequence. More recently, AlphaFold2 has been adapted for the prediction of multiple protein conformations by subsampling multiple sequence alignments (MSAs). The deterministic relationship between sequence and structure was hypothesized over half a century ago with profound implications for the biological sciences ever since. Based on this relationship, we hypothesize that protein conformational dynamics are also determined, at least in part, by amino acid sequence and that this relationship may be leveraged for construction of AI/ML models dedicated to predicting ensembles of protein structures (*i.e.*, distinct conformations). Accordingly, we conceptualized an AI/ML model architecture which may be trained on sequence data in combination with conformationally-sensitive structure data, coming primarily from nuclear magnetic resonance (NMR) spectroscopy. Sequence-informed prediction of protein structural dynamics has the potential to emerge as a transformative capability across the biological sciences, and its implementation could very well be on the horizon.

**Biological sequence information and its relationship with protein structure and dynamics**

The exchange of sequence information between biological macromolecules is a fundamental process of life. The pathway commonly summarized as DNA → RNA → protein, was put forward as the 'Sequence Hypothesis' by Francis Crick (Crick, 1958). The discovery of the double helix structure of DNA served as the initial inspiration, as asserted in one of scientific literature's most famous understatements: "It has not escaped our notice that the specific pairing we have postulated immediately suggests a possible copying mechanism for the genetic material" (Watson & Crick, 1953). On the relationship between sequence and structure, Crick further speculated that "folding is simply a function of the order of the amino acids" (Crick, 1958), and a more comprehensive articulation was later provided through the 'Thermodynamic Hypothesis' by Christian Anfinsen, which states that "the native conformation is determined by the totality of interatomic interactions and hence by the amino acid sequence" (Anfinsen, 1973). Protein structure is not static, however, and function may be dependent on conformational dynamics. For example, early studies on the structure of myoglobin revealed that structural re-arrangement is required for molecular oxygen binding (Kendrew et al., 1960; Miller & Phillips, 2021; Shulman et al., 1970). Building on these observations and hypotheses, we posit that 1D amino acid sequence determines both 3D structure and protein conformational dynamics **(Figure 1a)**. Together, the above insights span from information storage to biological function.

Looking back, it is quite remarkable that, despite the paucity of experimental evidence available at the time (Anfinsen, 1973; Crick, 1958), early hypotheses concerning the relationship between biological sequence and structure remain not only valid but central to major research advances (Cobb, 2017; Ille et al., 2022). Notably, the flow of biological information is distinct from the flow of energy and matter (Crick, 1958). While obeying the laws of chemistry and physics, biological information transfer and its deterministic role in structure may be treated as distinct and self-contained. Considering biological sequence information as such enabled significant progress in protein structure prediction. Jumper et al. emphasized the limitation of conventional physics-based approaches for this purpose and relied instead on sequence/structure-centric AI/ML approach to *de novo* protein structure prediction with AlphaFold2 (Jumper et al., 2021). This is not to say that the laws of chemistry and physics are not important to consider when dealing with the relationship between sequence and structure, but explicit implementation of chemistry/physics-based algorithms alone for protein structure prediction proved difficult, to say the least (Kryshtafovych et al., 2023). In contrast, utilization of AI/ML methods that combine

Protein Data Bank (PDB) holdings with genome sequence information to infer structure proved both feasible and remarkably effective (Baek et al., 2021; Jumper et al., 2021; Kryshtafovych et al., 2023; Lin et al., 2023).

Despite major progress with AI/ML approaches, application of sequence-structure relationships in this context is arguably in its infancy. AlphaFold2 and similar approaches make predictions of static protein structure, presumably an 'idealized' structure of a protein in a low energy state. However, proteins are not frozen in space and time—they are dynamic and adopt various structural conformations across complex energy landscapes (Frauenfelder et al., 1991; Henzler-Wildman & Kern, 2007; Miller & Phillips, 2021). Protein structural heterogeneity has been explored both experimentally and computationally. Nuclear magnetic resonance (NMR) spectroscopy has allowed for experimental determination of ensembles of conformational states (Alderson & Kay, 2021), while computational molecular dynamics (MD) simulations have aimed to provide conventional physics-based insights (Case et al., 2023; Hwang et al., 2024; Pall et al., 2020). Importantly, it should be emphasized that both NMR and MD simulations depend on physical properties of individual atoms and each approach offers a methodologically independent basis for characterizing relationships between amino acid sequence and conformational dynamics. Furthermore, combined incorporation of sequence and structural conformation data may be used for training of AI/ML models dedicated to sequence/structure-centric prediction of protein conformational ensembles.

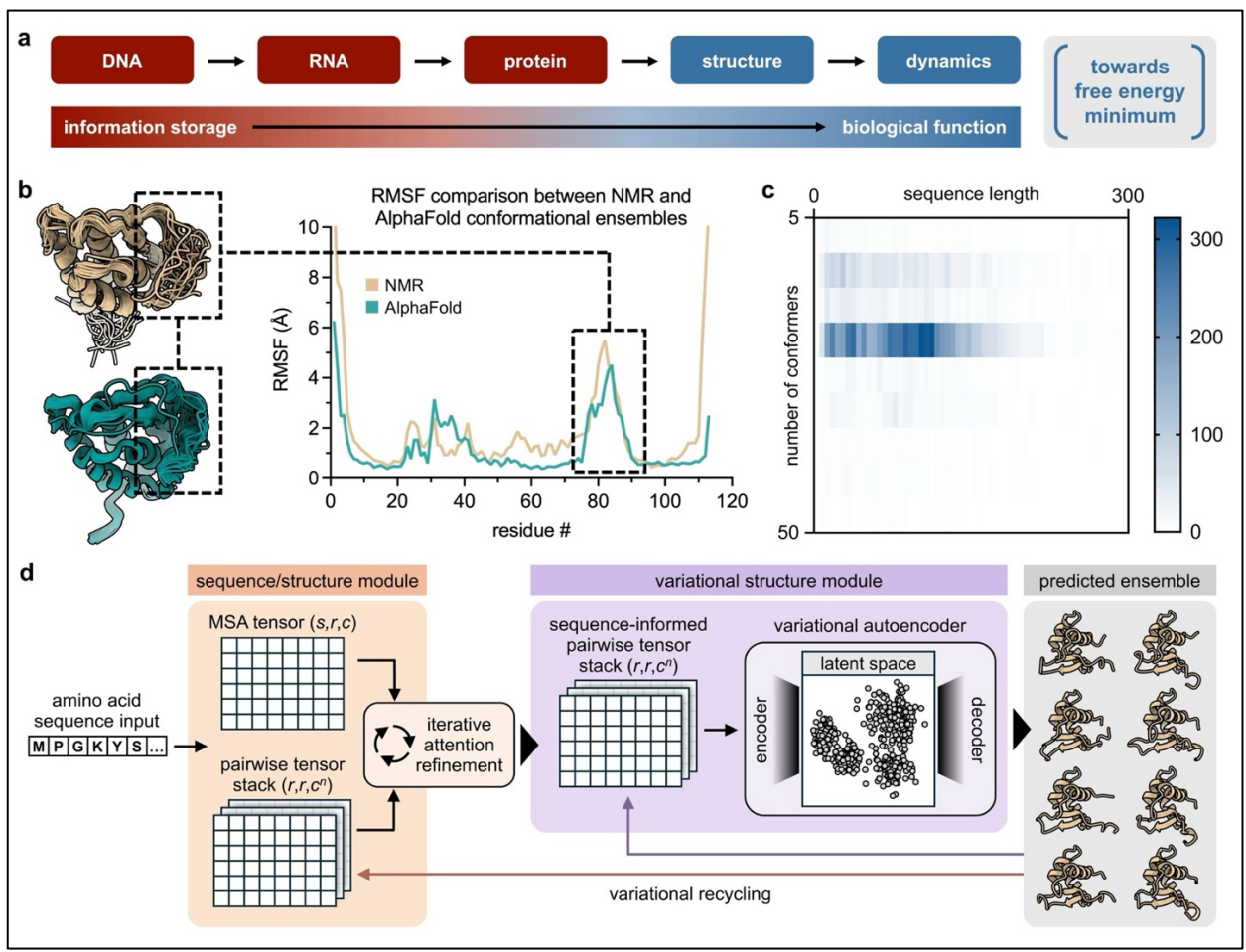

**Figure 1. Towards sequence/structure-centric prediction of protein conformational dynamics with AI/ML. (a)** Biological sequence information (red) encodes biophysical properties (blue) across a spectrum that spans from information storage to biological function. **(b)** NMR-determined (beige) (PDB ID 1GA3) and AlphaFold-predicted (teal) conformational ensembles of interleukin 13 along with a comparison of the root-mean-square fluctuation (RMSF) between them. The AlphaFold prediction was performed with stochastic MSA subsampling (Del Alamo et al., 2022; Kim et al., 2024; Monteiro da Silva et al., 2024; Wayment-Steele et al., 2024). **(c)** Distribution of NMR-determined single-chain protein entries currently deposited in the protein data bank (PDB) concerning number of conformational structures per ensemble (per entry) versus protein sequence length. The scale bar represents the number of PDB entries. **(d)** Conceptual AI/ML model architecture for end-to-end prediction of protein conformational ensembles from amino acid sequence input. The model comprises attention-based and variational mechanisms, representing a refined integration of existing models (Jumper et al., 2021; Mansoor et al., 2024). An NMR-determined conformational ensemble of the globular domain of human histone H1x (PBD ID 2LSO) is used for illustrative purposes in the predicted ensemble panel.

**Prediction of protein structure and dynamics with AI/ML**

Following the success of DeepMind in the Critical Assessment of Structure Prediction (CASP) competition (Kryshtafovych et al., 2021), the AI/ML model AlphaFold2 was recognized with the 2024 Nobel Prize in Chemistry for bridging the predictive gap between sequence and structure with unprecedented accuracy (Jumper et al., 2021; The Nobel Foundation, 2024). As previously mentioned, AlphaFold2 and similar AI/ML approaches, including RoseTTAFold (Baek et al., 2021) and ESMFold (Lin et al., 2023), rely on biological sequence data for prediction of protein structure via attention-based (Vaswani et al., 2017) neural network architectures. Three-dimensional (3D) atomic coordinate information from the PDB (Burley & Berman, 2021) was central for training these models. In addition, sequence data obtained from UniProt (UniProt Consortium, 2024) and other sources was also instrumental. AlphaFold2 and RoseTTAFold rely on multiple sequence alignments (MSAs), which carry information about co-evolution of pairs of amino acid residues, to inform 3D structure prediction. Remarkably, structure prediction accuracy arises in the early sequence-based stages of the AlphaFold2 model architecture (Jumper et al., 2021). Similarly, ESMFold, which is initially trained on sequence data alone using an attention-based language model without MSAs, exhibits substantial degradation of prediction accuracy when the sequence-based language model is dispensed with (Lin et al., 2023). The sequence/structure-centric frameworks of these AI/ML approaches rely heavily on the long-standing hypothesis that sequence determines 3D structure (Anfinsen, 1973; Crick, 1958). In the same vein, the predictive accuracy of these approaches bolsters support for this hypothesis—if sequence did not determine structure, AlphaFold2, and related approaches would not have been so successful.

More recently, AI/ML approaches have been introduced for prediction of multiple conformational states of proteins, rather than singular structures. AlphaFold2 was adapted to predict multiple protein conformations without retraining the model. More specifically, subsampling of sequences for assembly of MSAs has been demonstrated to result in predictions of multiple protein conformations that resemble conformations determined by experimental methods (Del Alamo et al., 2022; Monteiro da Silva et al., 2024; Wayment-Steele et al., 2024). Of particular note, AlphaFold2 was trained on structures determined by X-ray crystallography and cryogenic electron microscopy (cryoEM), but not more conformationally-sensitive spectroscopic methods. An example of a protein conformational ensemble predicted using this approach is shown in **Figure 1b**. Another AI/ML model developed by Mansoor et al., though not trained on biological sequence information, was able to predict multiple protein conformations with a variational autoencoder (VAE) architecture (Kingma & Welling, 2014; Mansoor et al., 2024). In this case, protein structures determined by X-ray crystallography accompanied by MD simulation snapshots were used to train the model to infer structural variation. 3D atomic coordinate-derived data is 'encoded' into a latent space of reduced complexity from which novel conformations are 'decoded' back into 3D structure data. While this model incorporates RoseTTAFold (Baek et al., 2021) to process the decoded structural information into 3D coordinate structures, sequence data were not used for training the underlying model (Mansoor et al., 2024). Nevertheless, the Mansoor et al. approach does provide a promising methodology for generating informative conformational ensembles from structural input.

Collectively, these approaches offer encouragement for development of AI/ML models dedicated to prediction of protein conformational dynamics in a sequence/structure-centric manner. Moreover, the combined incorporation of biological sequence data with conformationally-sensitive, experimentally-determined NMR data for model training is yet to be explored. There are over 10,000 single-chain protein 3D structure ensembles freely available from the PDB, which may be leveraged for training data **(Figure 1c)**. NMR data may be further enriched with carefully selected information from MD simulations. Furthermore, a sequence/structure-centric model may be developed without entirely re-inventing the wheel. We provide a conceptualization of the architecture of an AI/ML model dedicated to multi-conformational protein structure prediction **(Figure 1d)**, incorporating attributes from existing approaches including attention-based (Jumper et al., 2021) and VAE (Mansoor et al., 2024) mechanisms. The goal of such a model would be to input an amino acid sequence and perform end-to-end prediction of protein conformational ensembles, similar to prediction of static protein structures with AlphaFold2. Predicted 3D structure ensembles may be compared to NMR-determined ensembles for overall benchmarking of the method and individual accuracy assessments. The proposed approach represents a simplified conceptualization, and alternate approaches may be pursued by others. Whichever approaches are taken, the likelihood of success is strengthened by the capabilities of current AI/ML-based sequence/structure-centric models and recent forays into the prediction of multiple structural conformations, warranting further research and development.

**Conclusions**

Central to biology is the implicit relationship between amino acid sequence and 3D structure, as postulated nearly seven decades ago. This principle underpinned numerous discoveries and technical developments, including recent advances in AI/ML-based protein structure prediction. It appears highly likely that 1D sequence encodes not just a single idealized 3D structure but also the conformational dynamics of a protein and, therefore, biochemical/biological function. If this hypothesis holds, amino acid sequence data—alongside conformationally-sensitive structural data, *i.e.* NMR-determined structures—may be leveraged to train AI/ML models for prediction of conformational ensembles from amino acid sequence information alone. The functional implications of sequence-structure relationships across all of biology and biomedicine, literally spanning from agriculture to zoology (Burley et al., 2018) have been profound. This relationship will continue to play an important role at the nexus of AI/ML, data science, and biology.

**Acknowledgments**

Molecular graphics were prepared using UCSF ChimeraX, developed by the Resource for Biocomputing, Visualization, and Informatics at the University of California, San Francisco, with support from the National Institutes of Health R01-GM129325 and the Office of Cyber Infrastructure and Computational Biology, National Institute of Allergy and Infectious Diseases (Meng et al., 2023). The conformational ensemble of interleukin 13 in Figure 1b was predicted using the ColabFold implementation of AlphaFold2 (Kim et al., 2024). RCSB Protein Data Bank is jointly funded by the National Science Foundation (DBI-2321666, PI: S.K.B.), the US Department of Energy (DE-SC0019749, PI: S.K.B.), and the National Cancer Institute, the National Institute of Allergy and Infectious Diseases, and the National Institute of General Medical Sciences of the National Institutes of Health (R01GM157729, PI: S.K.B.).

**Author contributions**

A.M.I., E.A., M.B.M, and S.K.B. conceptualization; A.M.I. visualization; A.M.I. and E.A. writing–original draft; A.M.I., M.B.M., and S.K.B. writing–review & editing.

**Competing interests**

A.M.I. and E.A. are co-founders of North Horizon, which is engaged in the development of artificial intelligence-based software. E.A. is an employee of Microsoft. S.K.B. and M.B.M. declare no competing interests.